\documentclass[journal]{IEEEtran}
\usepackage[subpreambles=false]{standalone}
\usepackage{smoothing}
\usepackage{import}

\IEEEoverridecommandlockouts
\graphicspath{{img/}{sections/img/}}
\begin{document}
	\title{Unified Linearization-based Nonlinear Filtering}

	\author{
		\IEEEauthorblockN{Anton Kullberg\IEEEauthorrefmark{1}, Isaac Skog\IEEEauthorrefmark{2},~\IEEEmembership{Senior Member,~IEEE}, and Gustaf Hendeby\IEEEauthorrefmark{1},~\IEEEmembership{Senior Member,~IEEE}.
	\thanks{\noindent This work was partially supported by the Wallenberg AI,
		Autonomous Systems and Software Program (\textsc{WASP}) funded
		by the Knut and Alice Wallenberg Foundation.}
	}
  \IEEEauthorblockA{\IEEEauthorrefmark{1}%
    Dept.~Electrical Engineering, Linköping University, Linköping, Sweden}\\%
	\IEEEauthorblockA{\IEEEauthorrefmark{2}%
		Dept.~Electrical Engineering, Uppsala University, Uppsala, Sweden}\\%
	\IEEEauthorblockA{Email: \ttfamily \{anton.kullberg, gustaf.hendeby\}@liu.se, isaac.skog@angstrom.uu.se}
  }

\maketitle

\begin{abstract}
This letter shows that the following three classes of recursive state estimation filters: standard filters, such as the extended Kalman filter; iterated filters, such as the iterated unscented Kalman filter; and dynamically iterated filters, such as the dynamically iterated posterior linearization filters; can be unified in terms of a general algorithm. The general algorithm highlights the strong similarities between specific filtering algorithms in the three filter classes and facilitates an in-depth understanding of the pros and cons of the different filter classes and algorithms. We end with a numerical example showing the estimation accuracy differences between the three classes of filters when applied to a nonlinear localization problem.
\end{abstract}

\IEEEpubid{}


\section{Introduction}\label{sec:introduction}
\IEEEPARstart{S}{tate} estimation in nonlinear dynamical systems has been extensively studied in a wide variety of research fields.
Typical approaches employ some form of linearization-based approximate inference, which we focus on here.
These approaches linearize the nonlinear model locally (in each time instance) to employ the Kalman filter, which is the optimal estimator in the \emph{mean-squared error} (\abbrMSE) sense \cite{kalmanNewApproachLinear1960}.
Analytical linearization leads to the \emph{extended Kalman filter} (\abbrEKF), while sigma-point filters, such as the \emph{unscented Kalman filter} (\abbrUKF), \emph{cubature Kalman filter} (\abbrCKF), and similar can be thought of as statistical linearization filters \cite{julierNewApproachFiltering1995, kalmanNewApproachLinear1960, lefebvreCommentNewMethod2002}.
Statistical linearization filters also includes the Gaussian particle filter \cite{kotechaGaussianParticleFiltering2003,wuCommentsGaussianParticle2005}.



The estimation accuracy of linearization-based filters highly depend on the point (distribution in the statistical case) about which the models are linearized.
Typically, the linearization point (distribution) is chosen to be the mean (distribution) of the current state estimate.
With a large error in the state estimate, this can lead to compounding errors which, in the worst case, may cause the filter to diverge.
To alleviate this problem, several variants of iterated filters have been developed, such as the \emph{iterated extended Kalman filter} (\abbrIEKF), the \emph{iterated unscented Kalman filter} (\abbrIUKF), and the \emph{iterated posterior linearization filter} (\abbrIPLF) \cite{denhamSequentialEstimationWhen1965a, skoglundIterativeUnscentedKalman2019, zhanIteratedUnscentedKalman2007, sibleyIteratedSigmaPoint2006, garcia-fernandezPosteriorLinearizationFilter2015}.
These types of filters essentially iterate the measurement update, each time re-linearizing the measurement model with the \q{latest} iterate.
The efforts in iterated filtering have primarily been focused on finding a better linearization point for the measurement model, which has been motivated by the fact that nonlinearities in the measurement model affect the resulting state estimate to a greater extent than nonlinearities in the transition model.

Iterated filters have also been generalized to improve the linearization point for the transition model \cite{raitoharjuPosteriorLinearisationFilter2022,kullbergIteratedFiltersNonlinear2023}.
These algorithms, which we refer to as dynamically iterated filters, are essentially iterated one-step fixed-lag smoothers that extract information from the measurement at time $k$ to improve the linearization of the transition model at time $k-1$.
Examples of such algorithms are the \emph{dynamically iterated extended Kalman filter} (\abbrDIEKF), \emph{dynamically iterated unscented Kalman filter} (\abbrDIUKF), and the \emph{dynamically iterated posterior linearization filter} (\abbrDIPLF) \cite{raitoharjuPosteriorLinearisationFilter2022,kullbergIteratedFiltersNonlinear2023}.

In this letter, we seek to provide a \q{general} algorithm, from which all of the aforementioned filter algorithms can be derived as special cases.
In this way, we aim to clarify and highlight the strong similarities between different linearization-based filtering algorithms.
Thus, the contribution is a unification of linearization-based filters in a single general algorithm, encompassing analytically and statistically linearized, as well as iterated and non-iterated filters.
We also illustrate the performance differences between the three kinds of filter classes on an acoustic localization problem.

\section{Background}\label{sec:background}
For clarity, we here present analytical and statistical linearization within a common framework. The well-known Kalman filter and \emph{Rauch-Tung-Stribel} (\abbrRTS) smoother equations are also recapitulated.

\subsection{Kalman Smoother}
Assume an affine state-space model with additive Gaussian noise, of the form
\begin{subequations}
\begin{align}
\state_{k+1} &= \vec{A}_\dynmod\state_k + \vec{b}_\dynmod + \tilde{\pnoise}_k,& \tilde{\pnoise}_k&\sim\Ndist(\tilde{\pnoise}_k;\vec{0},\vec{Q}\!+\!\boldsymbol{\Omega}_\dynmod) \\
\obs_k &= \vec{A}_\obsmod\state_k + \vec{b}_\obsmod + \tilde{\onoise}_k, & \tilde{\onoise}_k&\sim\Ndist(\tilde{\onoise}_k;\vec{0},\vec{R}\!+\!\boldsymbol{\Omega}_\obsmod).
\end{align}
\end{subequations}
Here, $\state_k,~\obs_k,~\tilde{\pnoise}_k$ and $\tilde{\onoise}_k$ denote the state, the measurement, the process noise and the measurement noise at time $k$, respectively.
Lastly, assume that $\state_k\in \mathcal{X},\forall k$, where $\mathcal{X}$ is some set, typically $\mathbb{R}^{n_x}$, and that $\tilde{\pnoise}_k$ and $\tilde{\onoise}_k$ are mutually independent.
Note that usually, $\boldsymbol{\Omega}_\dynmod=\boldsymbol{\Omega}_\obsmod=\vec{0}$.
For this model, the (affine) Kalman smoother update equations are given by \cref{alg:kalmansmoother}, where subscript $_{k|k}$ denotes an estimate at time $k$ given measurements up until time $k$ and $K$ is the final time \cite{sarkkaBayesianFilteringSmoothing2013}.

\subsection{Analytical and Statistical Linearization}
Given a nonlinear model
\begin{equation*}
\vec{z} = \vec{g}(\state),
\end{equation*}
we wish to find an affine representation
\begin{equation}
\vec{g}(\state)\approx \vec{A}\state + \vec{b} + \eta,
\end{equation}
with $\eta\sim\Ndist(\eta; \vec{0}, \boldsymbol{\Omega})$.
In this affine representation, there are three free parameters, $\vec{A}, \vec{b}$, and $\boldsymbol{\Omega}$.
Analytical linearization through first-order Taylor expansion selects the parameters as
\begin{equation}\label{eq:analyticallinearization}
\vec{A}=\frac{d}{d\state}\vec{g}(\state)|_{\state=\bar{\state}},
\quad \vec{b} = \vec{g}(\state)|_{\state=\bar{\state}} - \vec{A}\bar{\state},
\quad \boldsymbol{\Omega}=\vec{0},
\end{equation}
where $\bar{\state}$ is the point about which the function $\vec{g}(\state)$ is linearized.
Note that $\boldsymbol{\Omega}=\vec{0}$ essentially implies that the linearization is assumed to be error free.

Statistical linearization instead linearizes w.r.t. a distribution $p(\state)$.
Assuming that $p(\state)=\Ndist(\state;\hat{\state},\vec{P})$, statistical linearization selects the affine parameters as
\begin{subequations}\label{eq:statisticallinearization}
\begin{align}
\vec{A} &= \Psi^\top \vec{P}^{-1}\\
\vec{b} &= \bar{\vec{z}}-\vec{A}\hat{\state}\\
\boldsymbol{\Omega} &= \Phi-\vec{A} \vec{P} \vec{A}^\top\\
\bar{\vec{z}} &= \mathbb{E}[\vec{g}(\state)]\\
\Psi &= \mathbb{E}[(\state-\hat{\state})(\vec{g}(\state) - \bar{\vec{z}})^\top]\\
\Phi &= \mathbb{E}[(\vec{g}(\state) - \bar{\vec{z}})(\vec{g}(\state) - \bar{\vec{z}})^\top],
\end{align}
\end{subequations}
where the expectations are taken w.r.t. $p(\state)$.
The major difference from analytical linearization is that ${\boldsymbol{\Omega}\neq 0}$, which implies that the error in the linearization is captured.

Typically, the expectations in \cref{eq:statisticallinearization} are not analytically tractable and thus, practically, one often resorts to some numerical integration technique.

\begin{algorithm}[tb]
    \caption{Kalman smoother}
    \label{alg:kalmansmoother}
    \begin{enumerate}
    \item Time update (TU)
    \begin{subequations}
    \label{eq:timeupdate}
    \begin{align}
    \hspace*{\dimexpr-\leftmargini}\hat{\state}_{k+1|k} &= \vec{A}_\dynmod\hat{\state}_{k|k} + \vec{b}_\dynmod\\
    \hspace*{\dimexpr-\leftmargini}\vec{P}_{k+1|k} &= \vec{A}_\dynmod\vec{P}_{k|k}\vec{A}_\dynmod^\top + \vec{Q} + \boldsymbol{\Omega}_\dynmod
    \end{align}
    \end{subequations}
    \item Measurement update (MU)
    \begin{subequations}
    \label{eq:measupdate}
    \begin{align}
    \hspace*{\dimexpr-\leftmargini}\hat{\state}_{k|k} &= \hat{\state}_{k|k-1} + \vec{K}_k(\obs_k - \vec{A}_\obsmod\hat{\state}_{k|k-1} - \vec{b}_\obsmod)\\
    \hspace*{\dimexpr-\leftmargini}\vec{P}_{k|k} &= \vec{P}_{k|k-1} - \vec{K}_k\vec{A}_\obsmod\vec{P}_{k|k-1}\\
    \hspace*{\dimexpr-\leftmargini}
    \vec{K}_k &\triangleq \vec{P}_{k|k-1}\vec{A}_\obsmod^\top(\vec{A}_\obsmod\vec{P}_{k|k-1}\vec{A}_\obsmod^\top + \vec{R} + \boldsymbol{\Omega}_\obsmod)^{-1}
    \end{align}
    \end{subequations}
    \item Smoothing step (S)
    \begin{subequations}
    \label{eq:smoothstep}
    \begin{flalign}
    \hspace*{\dimexpr-\leftmargini}\hat{\state}_{k|K} &= \hat{\state}_{k|k} + \vec{G}_k (\hat{\state}_{k+1|K} - \hat{\state}_{k+1|k})\\
    \hspace*{\dimexpr-\leftmargini}\vec{P}_{k|K} &= \vec{P}_{k|k} + \vec{G}_k(\vec{P}_{k+1|K} -\nonumber\\ &\quad~ \vec{A}_\dynmod\vec{P}_{k|k}\vec{A}_\dynmod^\top-\vec{Q}-\boldsymbol{\Omega}_\dynmod)\vec{G}_k^\top\\
    \hspace*{\dimexpr-\leftmargini}\vec{G}_k &\triangleq  \vec{P}_{k|k}\vec{A}_\dynmod^\top(\vec{A}_\dynmod\vec{P}_{k|k}\vec{A}_\dynmod^\top + \vec{Q} + \boldsymbol{\Omega}_\dynmod)^{-1}
    \end{flalign}
    \end{subequations}
    \end{enumerate}
\end{algorithm}

\section{Problem Formulation}\label{sec:problemformulation}
To set the stage for the unification of the different filter algorithms, the general state estimation problem is described here from a probabilistic viewpoint. 
To that end, consider a discrete-time state-space model (omitting a possible input $\inp_k$ for notational brevity) given by
\begin{subequations}\label{eq:ssm}
\begin{align}
\state_{k+1} &= \dynmod(\state_{k}) + \pnoise_{k}\label{eq:dyneq},& p(\pnoise_k) &= \Ndist(\pnoise_k;\vec{0}, \vec{Q})\\
\obs_k &= \obsmod(\state_k) + \onoise_k\label{eq:obseq},& p(\onoise_k) &= \Ndist(\onoise_k;\vec{0}, \vec{R}).
\end{align}
\end{subequations}
Note that \cref{eq:dyneq,eq:obseq} can equivalently be written as a \emph{transition density} and a \emph{measurement density} as
\begin{subequations}
\begin{align}
p(\state_{k+1}|\state_{k}) &= \Ndist(\state_{k+1};\dynmod(\state_{k}), \vec{Q})\label{eq:transitiondensity}\\
p(\obs_k|\state_k) &= \Ndist(\obs_k;\obsmod(\state_k), \vec{R})\label{eq:observationdensity}.
\end{align}
\end{subequations}
Further, the initial state distribution is assumed to be given by
\begin{equation}
p(\state_0)=\Ndist(\state_0;\hat{\state}_{0|0}, \vec{P}_{0|0}).
\end{equation}

Given the transition and measurement densities and a sequence of measurements $\obs_{1:k}=\{ \obs_i \}_{i=1}^k$, the filtering problem consists of computing the marginal posterior of the state at time $k$.
This can be done via the Bayesian recursions
\begin{subequations}\label{eq:statemarginaltimek}
    \begin{align}
        p(\state_k|\obs_{1:k-1}) &= \int_{\mathcal{X}}\! p(\state_k|\state_{k-1})p(\state_{k-1}|\obs_{1:k-1})d\state_{k-1}\label{eq:chapmankolmogorov}\\
        p(\state_k|\obs_{1:k}) &= \frac{p(\obs_k|\state_k)p(\state_k|\obs_{1:k-1})}{\vec{Z}_{k}}\\
        \vec{Z}_{k} &= \int_{\mathcal{X}} p(\obs_k|\state_k) p(\state_{k}|\obs_{1:k-1})d\state_k.
    \end{align}
\end{subequations}
In the case where $\dynmod$ and $\obsmod$ are linear, the (analytical) solution is given by the Kalman filter \cite{kalmanNewApproachLinear1960}.

In the general case, the marginal posteriors can not be computed analytically.
Inspecting \cref{eq:statemarginaltimek}, there are two integrals that require attention.
We turn first to the Chapman-Kolmogorov equation \cref{eq:chapmankolmogorov}.
Assuming that $p(\state_{k-1}|\obs_{1:k-1})$ is Gaussian, \cref{eq:chapmankolmogorov} has a closed form solution given by \cref{eq:timeupdate}, \emph{if}
$p(\state_k|\state_{k-1})$ is Gaussian and \cref{eq:dyneq} is affine.
Therefore, as \cref{eq:transitiondensity} is Gaussian, we seek an affine approximation of the transition function $\dynmod$ as
\begin{equation}\label{eq:dynapprox}
\dynmod(\state_{k-1}) \approx \vec{A}_\dynmod\state_{k-1} + \vec{b}_\dynmod + \eta_\dynmod,
\end{equation}
with $p(\eta_\dynmod) = \Ndist(\eta_\dynmod;\vec{0},\boldsymbol{\Omega}_\dynmod)$.
Hence, the transition density $p(\state_{k}|\state_{k-1})$ is approximated by $q(\state_{k}|\state_{k-1})$ as
\begin{equation}\label{eq:transdensityapprox}
q(\state_{k}|\state_{k-1}) = \Ndist(\state_{k}; \vec{A}_\dynmod\state_{k-1}+\vec{b}_\dynmod, \vec{Q}+\boldsymbol{\Omega}_\dynmod).
\end{equation}
If $\vec{A}_\dynmod,\vec{b}_\dynmod$, and $\boldsymbol{\Omega}_\dynmod$ are chosen to be the analytical linearization of $\dynmod$ about the mean of the posterior $p(\state_{k-1}|\obs_{1:k-1})$, the \abbrEKF time update is recovered through \cref{eq:timeupdate}.
Similarly, statistical linearization about $p(\state_{k-1}|\obs_{1:k-1})$ recovers the sigma-point filter time updates.
This yields an approximate predictive distribution $q(\state_k|\obs_{1:k-1})$, which can then be used to approximate the second integral of interest (and subsequently, the posterior at time $k$).
Explicitly, the second integral is approximated by
\begin{equation}\label{eq:normalizationapprox}
\vec{Z}_{k} \approx \int_{\mathcal{X}} p(\obs_k|\state_k)q(\state_k|\obs_{1:k-1})d\state_k.
\end{equation}
Similarly to \cref{eq:dynapprox}, \cref{eq:normalizationapprox} has a closed form solution \emph{if} $p(\obs_k|\state_k)$ is Gaussian and \cref{eq:obseq} is affine.
Thus, as \cref{eq:observationdensity} is Gaussian, we seek an affine approximation of the measurement function $\obsmod$ as
\begin{equation}\label{eq:obsapprox}
\obsmod(\state_k) \approx \vec{A}_\obsmod\state_k + \vec{b}_\obsmod + \eta_\obsmod,
\end{equation}
with $p(\eta_\obsmod) = \Ndist(\eta_\obsmod; \vec{0}, \boldsymbol{\Omega}_\obsmod)$.
Hence, the measurement density $p(\obs_k|\state_k)$ is approximated by $q(\obs_k|\state_k)$ as 
\begin{equation}\label{eq:obsdensityapprox}
q(\obs_k|\state_k) = \Ndist(\obs_k; \vec{A}_\obsmod\state_k + \vec{b}_\obsmod, \vec{R} + \boldsymbol{\Omega}_\obsmod),
\end{equation}
which leads to an analytically tractable integral.
With \cref{eq:transdensityapprox,eq:obsdensityapprox}, the (approximate) marginal posterior \cref{eq:statemarginaltimek} is now given by
\begin{equation}\label{eq:statemarginaltimekapprox}
q(\state_k|\obs_{1:k}) = \frac{q(\obs_k|\state_k)q(\state_k|\obs_{1:k-1})}{\int_{\mathcal{X}} q(\obs_k|\state_k)q(\state_k|\obs_{1:k-1})d\state_k},
\end{equation}
which is analytically tractable and given by \cref{eq:measupdate}.
Note that analytical linearization of \cref{eq:obsapprox} about the mean of $q(\state_{k}|\obs_{1:k-1})$ recovers the \abbrEKF measurement update, whereas statistical linearization recovers the sigma-point measurement update(s).

The quality of the approximate marginal posterior \cref{eq:statemarginaltimekapprox} directly depends on the quality of the approximations \cref{eq:transdensityapprox,eq:obsdensityapprox}.
The quality of \cref{eq:transdensityapprox,eq:obsdensityapprox} in turn directly depends on the choice of linearization points or densities, which is typically chosen to be the approximate predictive and previous approximate posterior distributions.
This choice is of course free and iterated filters, such as the \abbrIEKF, \abbrIUKF, and \abbrIPLF have been proposed to improve the approximation \cref{eq:obsdensityapprox} \cite{jazwinskiStochasticProcessesFiltering1970, garcia-fernandezIteratedStatisticalLinear2014, skoglundIterativeUnscentedKalman2019, zhanIteratedUnscentedKalman2007}.
These filters essentially iterate the measurement update to find an approximate posterior $q^i(\state_k|\obs_{1:k})$, which is used to re-linearize the function $\obsmod$ to produce a new approximation $q^{i+1}(\state_k|\obs_{1:k})$.
Iterated filters were recently generalized to dynamically iterated filters, which improve both the approximation \cref{eq:obsdensityapprox}, as well as the approximation \cref{eq:transdensityapprox} \cite{kullbergIteratedFiltersNonlinear2023,raitoharjuPosteriorLinearisationFilter2022}.
Dynamically iterated filters are essentially one-step iterated fixed-lag smoothers that produce both a better posterior approximation $q^{i+1}(\state_k|\obs_{1:k})$ as well as a smoothed approximation $q^{i+1}(\state_{k-1}|\obs_{1:k})$.

Next, we describe a unification of all of these filters in terms of one general algorithm, encompassing all possible variants of filters based on either analytical or statistical linearization.

\section{Unified Linearization-based Filtering}\label{sec:unification}
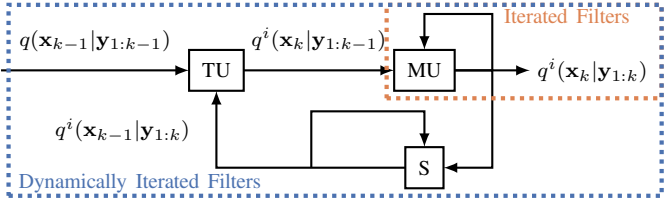
\begin{figure}[tb]
    \centering
    \begin{tikzpicture}[squarednode/.style={rectangle, draw=black, text opacity=1, thick, minimum size=5mm},
    every node/.style={inner sep=5pt,outer sep=0pt, font=\footnotesize}]
    \node[squarednode] (timeupdate) {TU};
    \node[squarednode] (measupdate) [right=of timeupdate, xshift=1cm] {MU};
    \node[squarednode] (smoothing) [below=of measupdate, yshift=.25cm] {S};
    \draw[-latex, thick] ($(timeupdate.west)+(-2.5, 0)$) -- node[label=above:{$q(\state_{k-1}|\obs_{1:k-1})$}, yshift=-.1cm] {} (timeupdate.west);
    \draw[-latex, thick] (timeupdate.east) -- node[label=above:{$q^i(\state_{k}|\obs_{1:k-1})$}, yshift=-.1cm] {} (measupdate.west);
    \draw[-latex, thick] (measupdate.east) -| ++(0.5, 0) |- (smoothing.east);
    \draw[-latex, thick] (measupdate.east) -- ++(1.0, 0) node[xshift=-.25cm, label=right:{$q^i(\state_{k}|\obs_{1:k})$}] {};
    \draw[-latex, thick] ($(measupdate.east)+(0.5, 0)$) -- ++(0, .75) -| (measupdate.north);
    \draw[-latex, thick] (smoothing.west) -| node[label=left:{$q^i(\state_{k-1}|\obs_{1:k})$}, yshift=.5cm] {} (timeupdate.south);
    \draw[-latex, thick] ($(smoothing.west)+(-1.25, 0)$) -- ++(0, .75) -| (smoothing.north);
    \draw[ultra thick, dotted, snsorange] (measupdate.north west)+(-.1, .6)
    node [xshift=2.4cm, yshift=-.6cm, label=above:{Iterated Filters}] {}
    rectangle ($(measupdate.south east)+(2.75, -0.1)$);
    \draw[ultra thick, dotted, snsblue] (timeupdate.north west)+(-2.4, .6)
    node [xshift=1.75cm, yshift=-1.9cm, label=below:{Dynamically Iterated Filters}] {}
    rectangle ($(smoothing.south east)+(2.9, -0.1)$);
    \end{tikzpicture}
    \caption{Schematic illustration of linearization-based filters. Iterated filters re-linearize the measurement update (MU). Dynamically iterated filters also re-linearize the time update (TU) through a smoothing step (S).}
    \label{fig:linfilter}
\end{figure}
We propose a unified linearization-based filtering algorithm that encapsulates a wide variety of existing algorithms.
The main idea behind the unification is that all linearization-based filters may be thought of as a single general algorithm, reduces to various special cases depending on specific implementation choices.
All of the filters are essentially centered around the three key steps \cref{eq:timeupdate,eq:measupdate,eq:smoothstep}.
They differ only in the choice of linearization strategy, as well as in which steps of the general (approximative affine) Kalman filter/smoother that are repeated or not.
The general algorithm is presented in \cref{alg:linbasedfilter} and encompasses standard linearization-based filters, iterated filters, and dynamically iterated filters. For clarity, it is also illustrated schematically in \cref{fig:linfilter}.
Note that the unified algorithm is purposefully restricted to algorithms that only require access to the latest measurement $\obs_k$, which, e.g., excludes the L-scan \abbrIPLF \cite{garcia-fernandezIteratedPosteriorLinearization2017}.

The linearization choices, which are assumed to be the same for all of the steps in the general filter algorithm, and the specific filter algorithms these choices lead to, are summarized in \cref{tab:algorithms}.
In \cref{tab:algorithms}, iterating either the measurement update (MU), both the time update (TU) and MU or none (--) is captured vertically, and the choice of particular linearization strategy horizontally.
Choosing analytical linearization inevitably leads to some form of \q{extended} version, i.e., either the \abbrEKF, \abbrIEKF, or \abbrDIEKF.
Statistical linearization is a bit more nuanced for two reasons. 
Firstly, it encapsulates a wide variety of algorithms, depending on the particular chosen statistical linearization, be it exact or approximated by, e.g., some form of cubature such as the \abbrCKF or \abbrUKF.
Note that we use \abbrCKF as a collective term for any statistically linearized Kalman filter based on sigma points, such as the smart sampling Kalman filter \cite{steinbringLRKFRevisitedSmart2014}, the spherical simplex-radial \abbrCKF \cite{wangSphericalSimplexRadialCubature2014}, or the multiple quadrate Kalman filter \cite{closasMultipleQuadratureKalman2012}.
Secondly, iterated versions of statistical linearization filters fall into two distinct categories, \abbrIUKF style that \q{freezes} the covariance update until the last iterate \cite{skoglundIterativeUnscentedKalman2019}, or \abbrIPLF style that continuously updates the covariance matrix -- essentially changing the sigma point spread each iteration \cite{garcia-fernandezIteratedStatisticalLinear2014}.
In \cref{tab:algorithms}, the \q{frozen} statistical linearization based filters are summarized by, e.g., the \abbrICKF and \abbrIUKF, but should be read as encapsulating any imaginable version of statistical linearization where the resulting filter has an update structure similar to that of the \abbrIEKF/\abbrDIEKF, i.e., with a \q{delayed} covariance update.
Note that the \q{freezing} or \q{delayed} behaviour of the \abbrIUKF/\abbrDIUKF is not explicitly defined in the algorithm \cref{alg:linbasedfilter} but amounts to setting $\vec{P}_{k-1|k}^{i+1}:=\vec{P}_{k-1|k}^i$ after \cref{alg:smoothstep} and $\vec{P}_{k|k}^{i+1}:=\vec{P}_{k|k}^i$ after \cref{alg:measupdate} until the last iteration.
\begin{table}[tb]
    \centering
    \captionsetup{justification=centering}
    \caption{Specific algorithms that \cref{alg:linbasedfilter} reduces to given different assumptions}
    \label{tab:algorithms}
    \resizebox{\columnwidth}{!}{
    \begin{tabular}{cccc}\toprule
        & \multicolumn{3}{c}{Linearization Method}\\\cmidrule{2-4}
        Iteration Type & Analytical & Statistical & Frozen Statistical\\ \midrule
        -- & \abbrEKF \cite{kalmanNewApproachLinear1960} & \abbrCKF\cite{arasaratnamCubatureKalmanFilters2009}, \abbrUKF\cite{julierNewApproachFiltering1995} & --\\
        MU & \abbrIEKF\cite{denhamSequentialEstimationWhen1965a} & \abbrIPLF \cite{garcia-fernandezIteratedStatisticalLinear2014} & \abbrICKF, \abbrIUKF\cite{sibleyIteratedSigmaPoint2006} \\
        TU \& MU & \abbrDIEKF\cite{kullbergIteratedFiltersNonlinear2023,wishnerEstimationStateNoisy1968} & \abbrDIPLF\cite{kullbergIteratedFiltersNonlinear2023,raitoharjuPosteriorLinearisationFilter2022} & \abbrDICKF, \abbrDIUKF\cite{kullbergIteratedFiltersNonlinear2023} \\ \bottomrule
    \end{tabular}}  
\end{table}

\begin{figure*}[t]
  \begin{minipage}{\columnwidth}
\begin{algorithm}[H]
  \caption{Linearization-based filter}
  \label{alg:linbasedfilter}
  \begin{algorithmic}[1]
  \Require $\hat{\state}_{k-1|k-1},\vec{P}_{k-1|k-1}, \obs_k$
  \State \cref{alg:timeupdate} with $\hat{\state}_{k-1|k}^0=\hat{\state}_{k-1|k-1}, \vec{P}_{k-1|k}^0=\vec{P}_{k-1|k-1}$
  \State \cref{alg:measupdate} with $\hat{\state}_{k|k}^0=\hat{\state}^1_{k|k-1}, \vec{P}_{k|k}^0=\vec{P}^1_{k|k-1}$
  \If{Standard Filter}\vspace{.25em}
    \State \textbf{return} $\hat{\state}_{k|k}, \vec{P}_{k|k}$\vspace{.25em}
  \ElsIf{Dynamically Iterated Filter}\vspace{.25em}
    \State \cref{alg:smoothstep}\vspace{.25em}
  \EndIf
  \State $i\gets 0$
  \While{not converged}\vspace{.25em}
    \If{Dynamically Iterated Filter}\vspace{.25em}
    \State Algs. \ref{alg:timeupdate}, \ref{alg:measupdate}, and \ref{alg:smoothstep}\vspace{.25em}
    \ElsIf{Iterated Filter} 
    \State \cref{alg:measupdate}\vspace{.25em}
    \EndIf
    \State $i\gets i+1$
  \EndWhile
  \State \textbf{return} $\hat{\state}_{k|k}^{i}, \vec{P}_{k|k}^{i}$ (and $\hat{\state}_{k-1|k}^{i}, \vec{P}_{k-1|k}^{i}$)
  \end{algorithmic}
\end{algorithm}
\end{minipage}
\begin{minipage}{\columnwidth}
  \begin{algorithm}[H]
      \caption{Linearized time update}
      \label{alg:timeupdate}
      \begin{algorithmic}[1]
      \Require $\hat{\state}_{k-1|k-1},\vec{P}_{k-1|k-1}, \hat{\state}_{k-1|k}^i, \vec{P}_{k-1|k}^i$
      \State 
        \parbox[t]{\dimexpr\linewidth-\algorithmicindent*2}{%
        Linearize $\dynmod$ about $\hat{\state}_{k-1|k}^i, \vec{P}_{k-1|k}^i$through \cref{eq:analyticallinearization} or \cref{eq:statisticallinearization}}
        \State Compute $\hat{\state}_{k|k-1}^{i+1}, \vec{P}_{k|k-1}^{i+1}$ by \cref{eq:timeupdate}
      \State \textbf{return} $\hat{\state}_{k|k-1}^{i+1},\vec{P}_{k|k-1}^{i+1}$
      \end{algorithmic}
    \end{algorithm}
  \vspace*{-\baselineskip}%
  
  \begin{algorithm}[H]
      \caption{Linearized measurement update}
      \label{alg:measupdate}
      \begin{algorithmic}[1]
      \Require $\hat{\state}_{k|k-1},\vec{P}_{k|k-1}, \hat{\state}_{k|k}^i, \vec{P}_{k|k}^i, \obs_k$
      \State 
      \parbox[t]{\dimexpr\linewidth-\algorithmicindent*2}{%
      Linearize $\obsmod$ about $\hat{\state}_{k|k}^i, \vec{P}_{k|k}^i$ through \cref{eq:analyticallinearization} or \cref{eq:statisticallinearization}}
      \State Compute $\hat{\state}_{k|k}^{i+1}, \vec{P}_{k|k}^{i+1}$ by \cref{eq:measupdate}\label{alg:line:mu2}
      \State \textbf{return} $\hat{\state}_{k|k}^{i+1},\vec{P}_{k|k}^{i+1}$
      \end{algorithmic}
    \end{algorithm}
  
  \vspace*{-\baselineskip}
  \begin{algorithm}[H]
      \caption{Linearized smoothing step}
      \label{alg:smoothstep}
      \begin{algorithmic}[1]
      \Require $\hat{\state}_{k-1|k-1},\vec{P}_{k-1|k-1}, \hat{\state}_{k|k},\vec{P}_{k|k}, \hat{\state}_{k-1|k}^i, \vec{P}_{k-1|k}^i$
      \State 
        \parbox[t]{\dimexpr\linewidth-\algorithmicindent*2}{%
        Linearize $\dynmod$ about $\hat{\state}_{k-1|k}^i, \vec{P}_{k-1|k}^i$through \cref{eq:analyticallinearization} or \cref{eq:statisticallinearization}}
        \State Compute $\hat{\state}_{k-1|k}^{i+1}, \vec{P}_{k-1|k}^{i+1}$ by \cref{eq:smoothstep}\label{alg:line:s}\vspace{.25em}
        \State \textbf{return} $\hat{\state}_{k-1|k}^{i+1},\vec{P}_{k-1|k}^{i+1}$
      \end{algorithmic}
    \end{algorithm}
  
  \end{minipage}
  \end{figure*}

\section{Numerical Example}\label{sec:example}
To demonstrate the application of the three types of filters, we consider a localization problem modeled by a nonlinear state-space model. To keep the results uncluttered, we only consider analytical linearization and focus our comparison on the \abbrEKF, \abbrIEKF, and \abbrDIEKF.

We consider a target maneuvering in a plane and describe the target state using the state vector $\state_k=\begin{bmatrix}p^x_k & v^x_k & p^y_k & v^y_k & \omega_k \end{bmatrix}^\top$. Here, $p^x_k,~p^y_k,~v^x_k$, and $v^y_k$ are the Cartesian coordinates and velocities of the target, respectively. Further, $\omega_k$ is the turn rate.
The transition model is given by
\begin{equation}
\state_{k+1} = \dynmod(\state_k) + \pnoise_k,
\end{equation}
where{\small
\begin{equation*}
\dynmod(\state_k) =
\begin{bmatrix}
1 & \frac{\sin(T \omega_k)}{\omega_k} & 0 & -\frac{(1-\cos(T \omega_k ))}{\omega_k} & 0\\
0 & \cos(T \omega_k ) & 0 & -\sin(T \omega_k ) & 0\\
0 & \frac{(1-\cos(T \omega_k ))}{\omega_k} & 1 & \frac{\sin(T \omega_k )}{\omega_k} & 0\\
0 & \sin(T \omega_k ) & 0 & \cos(T \omega_k ) & 0\\
0 & 0 & 0 & 0 & 1
\end{bmatrix}\state_k,
\end{equation*}
and $T$ is the sampling period. Further, $\pnoise_k \sim \Ndist(\pnoise_k;\vec{0},\vec{Q})$ is the process noise at time $k$, with
\begin{equation*}
\vec{Q} =
\mathrm{blkdiag}\left( 
\begin{bmatrix} q_1 \frac{T^3}{3} & q_1\frac{T^2}{2}\\
q_1\frac{T^2}{2} & q_1T \end{bmatrix},  
\begin{bmatrix} q_1 \frac{T^3}{3} & q_1\frac{T^2}{2}\\
q_1\frac{T^2}{2} & q_1T \end{bmatrix}, q_2 \right),
\end{equation*}}
where $q_1$ and $q_2$ are tunable parameters of the model.

The target emits a known sound pulse at a rate of $T=\SI{1.5}{\second}$ that is picked up by a set of four microphones.
With this, we construct time-difference-of-arrival (\abbrTDOA) observations, where each observation $i$ is modeled as
\begin{equation}
    \obs_k^i = r^1_k - r^i_k + \onoise_k,\quad i=1,\dots,3
\end{equation}
where $r^i_k\triangleq \lVert \begin{bmatrix}p^x_k & p^y_k\end{bmatrix}^\top - s^i \rVert$, and $s^i$ denotes the 2D position of the $i$:th microphone.
Further, $\onoise_k\sim\Ndist(\vec{0}, \vec{R})$, where $\vec{R}$ has been computed through a static calibration experiment.

We set $q_1=10^{-j},~q_2=10^{-l}$ and let $j=-6,\dots,0,~l=-5,\dots,0$, and sweep over all such pairs, i.e., 42 different process noise configurations.
For each configuration we compute the \abbrRMSE against a ground truth trajectory, obtained from a high-precision IR-marker positioning system.

\begin{figure}[t]
    \centering
    \includegraphics[width=\columnwidth]{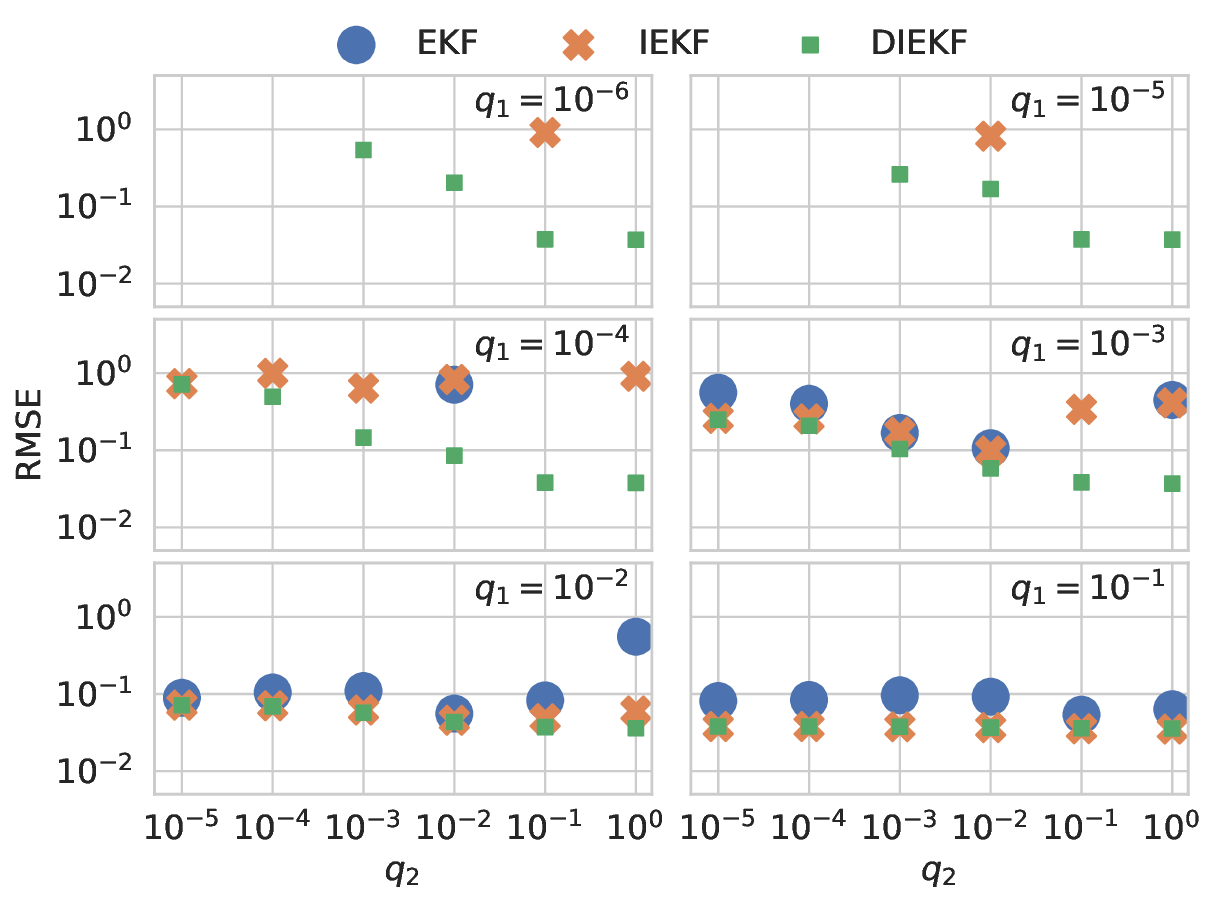}
    \caption{Positional \abbrRMSE for the \abbrEKF as blue dots, \abbrIEKF as orange crosses and \abbrDIEKF as green squares. Each subplot corresponds to a different value of $q_1$, indicated by the text in each subplot. An \abbrRMSE higher than approximately $\SI{1}{\meter}$ corresponds to a \q{divergent} filter based on visual inspection of resulting estimate trajectories and is left out of the plots.}
    \label{fig:rmse}
\end{figure}

The positional \abbrRMSE per noise configuration is presented in \cref{fig:rmse}.
Clearly, the \abbrDIEKF performs the best overall and is non-divergent in most cases. Here, divergence corresponds to an \abbrRMSE higher than $\SI{1}{\meter}$.
As the process noise is increased, the difference between the algorithms decreases, but the iterative procedure of the \abbrIEKF and \abbrDIEKF is still clearly beneficial.

\section{Conclusion}\label{sec:conclusion}
A unifying view of linearization--based nonlinear filtering algorithms has been presented.
It facilitates a comprehensive understanding of the commonalities and relationships between linearization--based standard, iterated, and dynamically iterated filters.
The presented algorithm is simple, easy to implement, and encompasses a wide range of existing filtering algorithms.
Lastly, the three classes of unified filtering algorithms were compared in a nonlinear localization problem, where the dynamically iterated filters were shown to be more resilient to poor process noise parameter tuning.

\bibliographystyle{IEEEtran}
\bibliography{ms}
\end{document}